\documentclass[12pt]{article}
\usepackage{latexsym}

\def\bs{\begin{subequations}}
\def\es{\end{subequations}}

\catcode`\@=11

\newtoks\@stequation
\def\subequations{\refstepcounter{equation}
  \edef\@savedequation{\the\c@equation}%
  \@stequation=\expandafter{\theequation}
  \edef\@savedtheequation{\the\@stequation}
  \edef\oldtheequation{\theequation}%
  \setcounter{equation}{0}%
  \def\theequation{\oldtheequation\alph{equation}}}

\def\endsubequations{\setcounter{equation}{\@savedequation}%
  \@stequation=\expandafter{\@savedtheequation}%
  \edef\theequation{\the\@stequation}\global\@ignoretrue}

\makeatletter
        \renewcommand{\theequation}{\thesection.\arabic{equation}}%
        \@addtoreset{equation}{section}%
\makeatother

\renewcommand{\thefootnote}{\fnsymbol{footnote}}

\begin{document}

\begin{titlepage}

Revised February 28, 2007  

(Main changes are in Sections 2 and 4.)

\begin{center}        \hfill   \\
            \hfill     \\
                                \hfill   \\

\vskip .25in

{\large \bf Relativistic Quaternionic Wave Equation II \\}

\vskip 0.3in

Charles Schwartz\footnote{E-mail: schwartz@physics.berkeley.edu}

\vskip 0.15in

{\em Department of Physics,
     University of California\\
     Berkeley, California 94720}
        
\end{center}

\vskip .3in

\vfill

\begin{abstract}
Further results are reported for the one-component quaternionic wave equation 
 recently introduced. A Lagrangian is found for the 
momentum-space version of the free equation; and another, nonlocal in 
time, is found for the complete equation. Further study of 
multi-particle systems has us looking into the mathematics of tensor 
products of Hilbert spaces. The principles of linearity and superposition  
are also clarified to good effect in advancing the quaternionic theory.
\end{abstract}


\vfill

\end{titlepage}

\renewcommand{\thefootnote}{\arabic{footnote}}
\setcounter{footnote}{0}
\renewcommand{\thepage}{\arabic{page}}
\setcounter{page}{1}

\section{Introduction}

In a recent paper \cite{CS}, we introduced the relativistic 
quaternionic wave equation, with $\psi = \psi(\textbf{x},t)$,
\begin{equation}
\frac{\partial}{\partial t} \psi\;i = \textbf{u}\cdot \nabla \psi + m 
\psi\;j = (i\frac{\partial}{\partial x} + j\frac{\partial}{\partial 
y} 
+ k\frac{\partial}{\partial z})\psi + m \psi\;j,\label{a1}
\end{equation}
which we call the free wave equation, and an extended version,
\begin{equation}
\frac{\partial}{\partial t} \psi\;i = \textbf{u}\cdot \nabla \psi +e\varphi \psi - 
e\textbf{u}\cdot\textbf{A} \psi\; i + m \psi\;e^{ieW}\;j,\label{a2}
\end{equation}
with external real potentials $\varphi, \textbf{A}, W$.  For either 
equation we found a conservation law, 
\begin{equation}\frac{\partial \rho}{\partial 
t} = \nabla \cdot \textbf{j},\label{a3}
\end{equation}
where 
\begin{equation}
\rho = \psi^{*}\psi, \;\;\;\;\; \textbf{j} = 
\{\frac{-i}{2},\psi^{*}\textbf{u}\psi\} \label{a4}
\end{equation}
and $\{X,Y\} = XY+YX$.

A central detail of this study is the recognition that we must 
allow numbers (quaternions) to multiply either on the left or on the 
right; so we use the notation $(a||b)\psi = a\psi b$.

Here we report some further results from this continuing study, 
including two versions of a Lagrangian, further exploration of 
many-particle states, and clarification of the principles of 
linearity and superposition. 

\section{A Lagrangian}

Formerly, we stated that we were unable to find a suitable Lagrangian 
for the quaternionic wave equation.  Here are  some new results. 

For the free wave equation, we go to momentum space:
\begin{equation}
\psi(\textbf{x},t) = \sum_{\textbf{p},\eta}\frac{1}{(2\pi)^{3/2}}\;exp(\eta \textbf{u}\cdot 
\hat{p}\;\textbf{p}\cdot \textbf{x}) \phi_{\textbf{p},\eta}(t),\label{b1}
\end{equation}
where $\hat{p} = \textbf{p}/p$, $p=|\textbf{p}|$,  and $\eta = \pm 1$.
Each component $\phi(t)$ (dropping the labels $\textbf{p},\eta$)  satisfies 
the equation 
\begin{equation}
\frac{d\phi}{dt} = \phi \Omega ,\label{b2}
\end{equation}
where $\Omega = i\eta p + km  = -\Omega^{*},\;|\Omega| = 
\sqrt{p^{2}+m^{2}}$,
and we find that the following Lagrangian
\begin{equation}
L = (\dot{\phi} - \phi \Omega)\phi^{*},\label{b3}
\end{equation}
when integrated over time, does provide a suitable action.  That is, 
variation of each of the four real components of $\phi$ does lead 
unambiguously to the specified equation of motion (\ref{b2}). 

This Lagrangian leads us to identify a new constant of the 
motion, in addition to $\phi\phi^{*}$, as follows:
\begin{equation}
\frac{d}{dt}\; \phi \Omega \phi^{*} = 0,\label{b4}
\end{equation}
and this looks like the quaternionic version (imaginary) of the Energy.
 
 This is unconventional in that we have multiplied the equation 
 (in $\phi$) by its conjugate $\phi^{*}$ on the \emph{right} rather 
 than on the left. (Alternatively, we can say that we should 
 re-identify which is the original wavefunction and which is its 
 complex conjugate.) We also note that this action is imaginary, not 
 real. Thus, there are actually 12 real equations resulting 
 from the variational principle; and they are all consistent,  
 implying  the four real equations originally given.

 In fact, if we write this Lagrangian as $L=iL_{1}+jL_{2}+kL_{3}$, 
any one of the real quantiies $L_{1}$ or $L_{2}$ or $L_{3}$ taken 
by itself is an adequate Lagrangian to recover the 
complete equation of motion (\ref{b2}) for the quaternionic amplitude 
$\phi$. 

Trying to find a coordinate space version of this Lagrangian, we are 
led to introduce the helicity projection operators,
\begin{eqnarray}
\pi_{\eta} = \frac{1}{2}[1-\eta \textbf{u}\cdot \nabla 
/\sqrt{-\nabla^{2}}], \;\;\;\;\; \eta = \pm 1 \\
\pi_{\eta}\;\pi_{\eta'} = \delta_{\eta,\eta'}\;\pi_{\eta}, \;\;\;\;\;
\textbf{u}\cdot \nabla\;\pi_{\eta} = -\eta \sqrt{-\nabla^{2}}\;\pi_{\eta}
\end{eqnarray}
and the helicity-projected wavefunctions, $\psi_{\eta} = 
\pi_{\eta}\psi(\textbf{x},t)$.  Then we can write the above Lagrangian as
\begin{equation}
L = \int d^{3}x\;\sum_{\eta} \;[\frac{\partial\;\psi_{\eta}}{\partial 
t} + \textbf{u}\cdot\nabla \psi_{\eta} i - m \psi_{\eta} k]\psi^{*}_{\eta}.
\end{equation}
This looks nice, but, of course, it is non-local in $\textbf{x}$; and I cannot 
extend this to 
include external potentials. See Section 4 for another attempt.

\section{Some Quaternion Identities}

Given an arbitrary quaternion $\psi$, there is the familiar  
identity $\psi^{*}\psi = \psi \psi^{*}$.  Here is something else:  
for any imaginary quaternion numbers $\alpha$ and $\beta$, we 
have the identity
\begin{equation}
\{\alpha,\psi^{*}\beta \psi\} = \{\beta, \psi\alpha \psi^{*}\}\label{c1}
\end{equation}
involving anticommutators $\{,\}$. The easiest way to prove this is 
to multiply the left hand side by $\psi \ldots \psi^{*}/|\psi|^{2}$. 
The result is the right hand side; but the left hand side is real, so 
it is unaffected by this operation.

For example, the conserved current $\textbf{j}$ which was previously written as 
$\{\frac{-i}{2},\psi^{*}\textbf{u}\psi\}$  can also be written as 
$\{\frac{-\textbf{u}}{2},\psi i\psi^{*}\}$.

Writing $\psi = \psi_{1}+\psi_{2}$, we have two further identities:
\begin{eqnarray}
\psi_{1}^{*}\psi_{2} + \psi_{2}^{*}\psi_{1} = 
\psi_{2}\psi_{1}^{*} + \psi_{1}\psi_{2}^{*} \label{c2}\\
\{\alpha,\psi_{1}^{*}\beta \psi_{2}+\psi_{2}^{*}\beta \psi_{1}\} =
\{\beta, \psi_{2}\alpha \psi_{1}^{*}+\psi_{1}\alpha 
\psi_{2}^{*}\}.\label{c3}
\end{eqnarray}

In a similar vein we can generalize the conservation law (\ref{a3})   
to include a mixed density $\rho_{1,2} = \psi_{1}^{*}\psi_{2} + 
\psi_{2}^{*}\psi_{1}$, with a similar expression for the current 
$\textbf{j}_{1,2}$.

A special case of the  result (\ref{c3}), following the 
substitution $\psi_{1} \rightarrow \beta \psi_{1}$,  is
\begin{equation}
\{\alpha, \psi_{1}^{*} \psi_{2}-\psi_{2}^{*} \psi_{1}\} = 
2(\psi_{2}\alpha \psi^{*}_{1} - \psi_{1}\alpha \psi^{*}_{2}).\label{c4}
\end{equation}
Using the above formulas, we can rewrite the momentum space Lagrangian of 
Section 2 as follows:
\begin{equation}
L_{1} = \{\frac{-i}{2},L\} = -\phi^{*}\;i\;\dot{\phi} + \frac{1}{2}
\{\Omega,\phi^{*}\;i\;\phi\}.\label{c5}
\end{equation}

\section {Another Lagrangian}

We have been able to find a Lagrangian for the extended wave 
equation, but it is non-local in time.

Define
\begin{eqnarray}
\psi = \psi(\textbf{x},t), \;\;\;\;\; \bar{\psi} = 
\psi^{*}(\textbf{x},-t) \label{d1}\\
\epsilon = \epsilon (t) = +1 (t > 0),\;\;or\;\;-1 (t<0)\label{d2}
\end{eqnarray}
and assume that the potential $\varphi$ is even under $t\rightarrow 
-t$ while $\textbf{A}$ and $W$ are odd.

Then we  construct the following (real) action,
\begin{equation}
\mathcal{A}= \int d^{4}x\;[\bar{\psi}\;\frac{\partial \psi}{\partial t} 
+\{\frac{i}{2},\epsilon \bar{\psi}( \textbf{u}\cdot \nabla   + 
e\varphi) \psi\} + e \bar{\psi}\textbf{u}\cdot 
\textbf{A} \psi  - \frac{m}{2} \{e^{i\epsilon eW}k,\epsilon\bar{\psi}\psi\} ],
\label{d3}
\end{equation}
which, upon variation of the four real components of $\psi$, gives 
the extended wave equation as follows:
\begin{equation}
\epsilon\frac{\partial \psi}{\partial t} \;i = \textbf{u}\cdot \nabla \psi +e\varphi \psi - 
\epsilon e\textbf{u}\cdot\textbf{A} \psi\; i + m \psi\;e^{i\epsilon 
eW}\;j.\label{d4}
\end{equation}
For $t > 0$ this is exactly the original equation (\ref{a2}); and if 
we now change $t \rightarrow -t$, we find the very same equation 
for $\psi(\textbf{x},-t)$.

 One may well wonder what is the correct choice of time $t=0$ and how 
 to interpret the discontinuity which this Lagrangian posits at that point.

\section{Multi-Particle States}

In Reference \cite{CS}, Section 12,  we saw a particular way to construct two-or-more-particle 
wavefunctions for a generalization of our quaternionic wave equation; 
and that will be explored further below.

In response to some comments received, I did make an attempt to expand 
the established mathematics 
of ``tensor products'', as applied to  Hilbert spaces over an Abelian 
field (like ordinary complex numbers), so that it might accommodate 
quaternionic quantum theory. This effort is summarized in 
Appendix A; and it may be called a limited success.  

Let me start by repeating the previous approach used, with a slight 
difference of notation, which I will explain a bit later on. Here, we 
will make frequent use of the notation for two-sided multiplication 
by quaternionic numbers and functions: $(a||b)\psi = a\psi b$.

Write the free wave equation, Eq. (\ref{a1}), as
\begin{equation}
\mathcal{D} \psi = [(\frac{\partial}{\partial t}||k) - (\textbf{u} 
\cdot \nabla ||j) + m] \psi = 0,\label{e1}
\end{equation}
with our regular plane wave solution written as 
\begin{equation}
\psi(\textbf{x},t) = \psi^{op}_{\textbf{p},\eta}(\textbf{x},t) \phi = 
(\;\exp(\eta \textbf{u} \cdot \hat{p} \textbf{p}\cdot \textbf{x}) || 
exp((i\eta p + km)t)\;)\phi.\label{e2}
\end{equation}
With $x$ standing for the four coordinates $\textbf{x},t$, the propagator is written as 
\begin{equation}
G_{+}(x,x') = -(\theta (t-t')||k)\sum_{\textbf{p},\eta}\frac{1}{(2\pi)^{3}}\;
\psi^{op}_{\textbf{p},\eta}(\textbf{x}-\textbf{x}',t-t'),\label{e3}
\end{equation}
so that we have 
\begin{equation}
\mathcal{D}\;G_{+} = \delta^{4}(x-x').\label{e4}
\end{equation}

Two of these definitions, Eqs. (\ref{e1}) and (\ref{e3}), differ 
from what was given before by the extra quaternion $k$ seen multiplying 
from the right. The reason for this becomes apparent when we note how 
the Lorentz transformation is applied to the wave equation:
\begin{equation}
\mathcal{D} \rightarrow \mathcal{L}^{-1}\mathcal{D}\mathcal{L}, 
\;\;\;\;\; \mathcal{L} = exp((\frac{1}{2}\textbf{u}\cdot \textbf{v} 
||i)).\label{e5}
\end{equation}
This is important for maintaining Lorentz covariance in some other steps, 
which involved products of these $\mathcal{D}$ operators. 

If we can read Eq. (\ref{e4}) as saying $ \mathcal{D} G_{+} = 1$, can 
we also have $G_{+}\mathcal{D} = 1$? We investigate:
\begin{eqnarray}
G_{+}\;\mathcal{D}\;\psi(\textbf{x},t) = -\int d^{4}x' 
(\theta(t-t')||k)\sum_{\textbf{p},\eta}
\frac{1}{(2\pi)^{3}} ( \;exp(\eta \textbf{u}\cdot 
\hat{p}\;\textbf{p}\cdot 
(\textbf{x}-\textbf{x}')) \nonumber \\ ||  
exp(\Omega_{p,\eta}(t-t')) \;) [(\frac{\partial}
{\partial t'}||k) - (\textbf{u} \cdot \nabla ' ||j) + 
m]\;\psi(\textbf{x}',t').
\end{eqnarray}
If we execute partial integration on the (primed) space and time 
derivatives,  
we do get exactly the answer $\psi(\textbf{x},t)$, provided 
that we understand one detail of the mathematical notation here. The 
symbol $||$ that stands to the left of the $(t-t')$ argument in the 
Green function does NOT mean that these coordinates are to be placed 
to the right of the time derivative operator which follows.  That 
``right-multiplication'' instruction applies only to numbers 
(quaternions) and not to coordinates.  Our mathematical notation may 
need to be improved somewhat to make this rule transparent; but we 
shall deal with what we now have. 

Let us now see what sense we can make of the most general two-particle 
wavefunction, for the free particles, written in our ``nested'' way:
\begin{eqnarray}
\Psi(\textbf{x}_{1},t_{1};\textbf{x}_{2},t_{2}) =
\sum_{\textbf{p},\eta}\;exp(\eta \textbf{u}\cdot 
\hat{p}\;\textbf{p}\cdot\textbf{x}_{1})\;\psi_{\textbf{p},\eta}(\textbf{x}_{2},t_{2})
\; exp((i\eta p + km)t_{1})\label{e6} \\
\psi_{\textbf{p},\eta}(\textbf{x}_{2},t_{2}) = \sum_{\textbf{p}',\eta'}\;exp(\eta' \textbf{u}\cdot 
\hat{p'}\;\textbf{p}'\cdot\textbf{x}_{2})\;\phi_{\textbf{p},\eta,\textbf{p}',\eta'}
\; exp((i\eta' p' + km)t_{2}).\label{e7}
\end{eqnarray}
Let us now study $\rho = \Psi^{*}\Psi$, which we understand to be a 
function of two sets of space-time coordinates. A detailed 
calculation, which I will not write out here, leads to the 
nice result
\begin{equation}
\frac{\partial \rho}{\partial t_{1}} = \nabla_{x_{1}} \cdot 
\textbf{j}, \;\;\;\;\; \textbf{j} = 
\{\frac{-i}{2},\Psi^{*}\textbf{u}\Psi\}.\label{e8}
\end{equation}
This looks exactly like the one-particle conservation law (\ref{a3}); but it has 
a broader interpretation now. If we identify $\rho$ as a probability density, then 
this is a joint probability density that is conserved - looking at 
coordinates of particle number 1 - for any given  wavefunction of 
particle number 2. Such a result in ordinary (complex) quantum theory 
would be rather obvious and would be interpreted in terms of the 
factorizeability of the theory for two non-interacting particles.

For the quaternionic theory, this result is not trivial, especially 
in view of the previous difficulties encountered by other authors: 
the so-called failure of clustering. (See Reference \cite{SA}).

This does not solve all such problems, however, since our starting 
wavefunction (\ref{e6}) has the coordinates 1 and 2 ordered in a 
particular way.  The result (\ref{e8}) does not hold true for 
coordinates 2 in the derivative operations. However, if we first 
integrate over coordinate $\textbf{x}_{1}$, then we do recover 
a one-particle density which is conserved.

\section{More on Quaternionic Amplitude and Linearity}

In Reference \cite{CS}, Section 7 , we noted that the amplitude constant $\phi$ of a plane wave solution of 
the free equation stood in a particular place in the middle of the
complete wavefunction, 
\begin{equation}
\psi = exp(\eta \textbf{u}\cdot \hat{p}\;\textbf{p}\cdot 
\textbf{x})\; \phi \; exp((i\eta p + km)t),\label{f1}
\end{equation}
and not casually at the right or the left of the wavefunction as one 
might place it for convenience in ordinary (complex) quantum theory.
One may ask whether this feature is generally true, even for the full 
interacting wave equation.

Let us write the most general quaternionic wave equation as
$\frac{\partial \psi}{\partial t} = H \psi$. Then the general 
initial value problem can be resolved, formally at least, as 
\begin{equation}
\psi(t) = exp(Ht)\;\psi(0)\label{f2}
\end{equation}
where $\psi(0)$  
may be specified arbitrarily and we still have a proper solution of 
the wave equation. Note that we have suppressed the space 
coordinates, so this initial data $\psi(0)$ can really be any function of 
space. Also, the operator $H$ is likely to involve 
``right-multiplying'' quaternions, so this $\psi(0)$ really stands in 
the midst of a complicated set of other things, just like the 
constant $\phi$ in (\ref{f1}). If the operator $H$ should contain time dependent terms, we 
also know how to rewrite the propagator as a time-ordered exponential 
of the integral $\int_{0} ^{t}\;H(t')dt'$.

This initial quaternionic data $\psi(0)$ generalizes the simple 
constant $\phi$ which we called the amplitude of the plane wave; and 
from this identification we see the generalized meaning of linearity 
and superposition for quaternionic quantum theory.

Let us further scrutinize the fundamental meaning of the principle of linearity 
(or superposition), which is so central to quantum theory.  In 
the usual (complex) theory, we say that if $\psi_{1}$ and $\psi_{2}$ 
are solutions of the wave equation (or state vectors in the Hilbert 
space), then
\begin{equation}
\psi = c_{1}\psi_{1} + c_{2}\psi_{2},\label{f3}
\end{equation}
for arbitrary complex numbers $c_{1},c_{2}$, shall also be a solution of the 
wave equation (or a state vector). When this definition of linearity is carried 
over into quaternionic theory, it causes much trouble (see Appendix 
A, for example), even aside from the question of whether to write 
those constants on the left or on the right.

Let us ask what is really required in physics by examining the familiar example 
of the two-slit interference experiment, which is discussed in any 
textbook.  Let $\psi_{1}(x_{1})$ be the amplitude in the open slit number 
1; and 
let $\psi_{2}(x_{2})$ be the amplitude in the open slit number 2.  Then we want 
to see how each of those wavefunctions propagates to some point $x$ on 
the observation screen:
\begin{equation}
\psi_{1}(x) = G(x;x_{1})\psi_{1}(x_{1}), \;\;\;\;\; \psi_{2}(x) = 
G(x;x_{2})\psi_{2}(x_{2})\label{f4}
\end{equation}
where the propagators $G(x,x')$, are derived from the pertinent wave 
equation.  Finally, the principle of superposition tells us merely 
that we should add these two amplitudes at the observation screen:
\begin{equation}
\psi(x) = \psi_{1}(x) + \psi_{2}(x).\label{f5}
\end{equation}
Whatever phase difference arises between these two waves (thus 
producing the interference pattern) comes from the propagators and it 
is thus already built into the correct wavefunctions at point $x$.

The mathematical statement of superposition (\ref{f5}) is much simpler than the 
statement (\ref{f3}); and, I claim,  this is all we need for an 
acceptable quantum theory.  This makes a big difference in any investigation of 
the mathematics of 
quaternionic quantum theory.

\vskip 1.0cm
\begin{center} \textbf{ACKNOWLEDGMENTS} \end{center}

 I am grateful to S. Adler, W. Arveson, K. Bardakci and M. Rieffel 
 for some helpful conversations.

\vskip 1.5cm
\setcounter{equation}{0}
\def\theequation{A.\arabic{equation}}
\boldmath
\noindent{\bf Appendix A: Tensor Products}
\unboldmath
\vskip 0.5cm

A long-standing problem in earlier studies of quaternionic quantum 
theory has been the description of multi-particle states.
First I will review how we do this in the usual (complex) quantum 
theory and then explore how the usual mathematics of tensor products 
may be extended to the realm of quaternions.   We start with two 
particles, given coordinates $x_{1}$ and $x_{2}$.

The usual tensor product formalism has us write a composite vector as
\begin{equation}
\psi(x_{1},x_{2}) = f(x_{1})g(x_{2}) \rightarrow |\psi> = f \otimes g 
\label{A1}
\end{equation}
This composite is supposed to reside in a new Hilbert space 
with the inner product rule
\begin{equation}
<\psi'|\psi> = <f'|f><g'|g>\label{A2}
\end{equation}
which we say is \emph{factorizeable}. We also note the complete 
\emph{linearity} of this inner product under $f \rightarrow 
c_{1}\;f_{1} + c_{2}\;f_{2}$ for arbitrary complex numbers 
$c_{1},c_{2}$; and similarly for $g,f',g'$.
Furthermore, we have composite operators that act on such states as
\begin{equation}
(A \otimes B)  (f\otimes g) = (Af)\otimes(Bg),\label{A3}
\end{equation}
which property we may call \emph{separability} of operators.

In the complex case, any numbers that occur as multipliers in the 
operators or in the state vectors can be factored out and written 
anyplace we wish, since they commute with all other operators and 
with one another.  In the quaternionic case that simplicity no longer 
holds. In earlier study of this tensor product mathematics, Horwitz 
and Biedenharn \cite{HB} concluded that it was impossible to extend 
this to quaternionic quantum mechanics.  Their primary criterion was 
the \emph{linearity} condition, which we see obviously fails when we 
have quaternionic functions $f$ and $g$. This same conclusion is given 
in Adler's book \cite{SA}.

However, as discussed above, we now have a more enlightened view of 
linearity and superposition in dealing with quaternionic 
wavefunctions. So we shall require only the simple sort of linearlty: 
$f \rightarrow f_{1}+f_{2}$, which also allows for real 
multipliers.

This still leaves us with plenty of other criteria to be checked out; 
and this is what we shall do here.

The first place we see trouble in trying to extend the above 
formalism to quaternions is Eq. (\ref{A2}), since the two 
single-particle inner products do not commute with one another and 
that violates the general rule in any Hilbert space
\begin{equation}
<\psi'|\psi>^{*} =<\psi|\psi'>.\label{A4}
\end{equation}

\vskip 1cm 

$\;\;\;\;\;$FIRST ATTEMPT

Let us now try an alternative rule for inner products,
\begin{equation}
<\psi'|\psi> = \frac{1}{2}[<f'|f><g'|g>+<g'|g><f'|f>].\label{A5}
\end{equation}
which at least satisfies Eq. (\ref{A4}). But there is more to check.
Consider the particular composite vector
\begin{equation}
|\Psi_{o}> = f \otimes g + (fi)\otimes (gi)
\end{equation}
which, using the rule (\ref{A5}), turns out to have norm 
$<\Psi_{o}|\Psi_{o}> = 0$. Yet, when we take the inner product 
$<f'\otimes g'|\Psi_{o}>$, we find that it is nonzero for general 
quaternionic functions $f$ and $g$.  This is an intolerable 
situation; and so we reject this rule (\ref{A5}).

\vskip 1cm 

$\;\;\;\;\;$SECOND ATTEMPT

Alternatively, we consider the rule 
\begin{equation}
<\psi'|\psi> = <g'|<f'|f>|g>,\label{A6}
\end{equation}
which involves one number (the inner product $<f'|f>$) sitting inside 
an inner product of two other vectors, $<g'|\ldots|g>$. For $f'=f, 
g'=g$ this is simply $<\psi|\psi>=<f|f><g|g>$.

Searching for a zero norm composite vector we now come up with
\begin{equation}
|\Psi_{o}> = f \otimes g + (fi) \otimes (ig).\label{A7}
\end{equation}
Of course, for simply complex functions it is perfectly obvious that 
this is just $[1-1]f\otimes g$. Is this a problem, however, 
for our quaternionic situation? We do not see the same problem as 
found above, in the first attempt, since, according to the rule (\ref{A6}), the 
general inner product $<f' \otimes g'|\Psi_{o}> = 0$.

This suggests a mathematical scheme for our tensor products 
 where it is stated that $fa\otimes g = f \otimes ag$ for any scalar 
 a. This
looks like the ordinary rule for associativity in multiplication; 
however, this does cause us a problem with the idea of 
\emph{separability}.  

We have said that we want to have
$(A\otimes B)(f\otimes g) = 
(Af)\otimes(Bg)$, which is to say that we can assign a quaternionic 
operator to a specific subspace.  But we  have now acknowledged that 
$(fa)\otimes g = f \otimes (ag)$. These two properties conflict, as may 
be seen in 
\begin{eqnarray}
(I\otimes B) ((fa)\otimes g) = (fa)\otimes (Bg)= f \otimes (aBg) \\
(I\otimes B) ((fa)\otimes g)=(I\otimes B) (f\otimes(ag))=
f \otimes (Bag).\label{A8}
\end{eqnarray}
This is consistent only if $a$ always commutes with $B$, which our 
quaternions do not admit.  So this second attempt 
is also a failure.

\vskip 1cm

$\;\;\;\;\;$THIRD ATTEMPT

As a variant of the second attempt, we consider the inner product rule 
\begin{equation}
<\psi'|\psi> = \frac{1}{2}[<g'|<f'|f>|g> + <f'|<g'|g>|f>],\label{A9}
\end{equation}
which asserts that the order of the components $f,g$ should not 
matter. Again, take a general superposition $|\Psi> = f \otimes g + 
f'\otimes g'$ and calculate the norm, which can be rearranged as 
follows.
\begin{eqnarray}
N^{2} \equiv <f|f><g|g>,\;\; N'^{2}\equiv <f'|f'><g'|g'>, \;\; M \equiv 
NN' \nonumber \\
X \equiv \frac{1}{M}<g'|<f'|f>|g>, \;\;\;\; Y \equiv 
\frac{1}{M}<f'|<g'|g>|f> \nonumber \\
<\Psi|\Psi> = (N-N')^{2} + \;\;\;\;\;\;\;\;\;\;\;\;\;\;\;\;\;\;\;\; \label{A10}\\
\frac{M}{2}\left[ |1+X|^{2} +  |1+Y|^{2} 
+ (1-|X|^{2}) + (1-|Y|^{2}) \right] .\nonumber 
\end{eqnarray}
 Each of the five terms in 
Eq. (\ref{A10}) is non-negative, so for this norm to vanish, 
each of those five terms must vanish. The  $X$ terms in (\ref{A10}) 
will vanish if we choose $f'=f\mu$ and $g'=\nu g$ with $\mu\;\nu = 
-1$. This leads to the troublesome situation we found above  in the 
second attempt. Now, however,
 we also need the  $Y$ terms in (\ref{A10}) to vanish; 
and these two requirements can be met, for general 
quaternionic functions $f,g$, only with the 
choices $f'=f,\;g'=-g$ or $f'=-f,\;g'=g$, which gives us trivially, 
$|\Psi>=0$.

This inner product rule (\ref{A9}) is unfamiliar; it says, for 
example, that the vectors $1\otimes 1$ and $i \otimes j$ are 
orthogonal to each other. Nevertheless, the mathematics appears to be 
consistent.

So, it appears that we have now overcome some of the previous difficulties; 
and with this last choice for an inner product rule, (\ref{A9}), we can 
indeed achieve both the enlightened version of \emph{linearity} as 
well as the 
attractive idea of \emph{separability}  given by (\ref{A3}), both 
within the confines of a proper Hilbert space.  I do not yet have a 
complete proof that this 
is correct; but I am encouraged by the   idea that what we 
required is merely a decent method of bookkeeping. We can keep the 
individual Hilbert spaces effectively separated but still have a proper composite 
Hilbert space with this  tensor product.

Thus, if we have two operators, $A$ acting on the vector $f$ and $B$ 
acting on the vector $g$, then we can write them, in the tensor 
product notation, as $A \otimes I$ and $I \otimes B$, respectively; 
and these two operators now commute with each other regardless of any 
quaternions embedded in them.

How to extend this rule from two to n components in the tensor product is an open 
question.  It might stay at two terms in the inner product: a given 
order and its reverse. Or it might become $2^{n-1}$ terms by a process 
of induction; or it might become n!, with all permutations of ordering.

What about the criterion of \emph{factorizeability}? We have 
clearly had to give up on that.  I think this is more a matter of 
physics than of mathematics.  If this quaternionic theory has any relation to 
physical reality, then it appears that we shall have to acknowledge 
some sort of  
persistent entanglement of many-particle states. That is not nice, 
according to our conventional ideas and experience; but it is not 
unimaginable. The example of two-particle density studied in Section 
5, above, gives us a mixed picture of such entanglement.

\end{document}